\documentclass[prl,twocolumn,amsmath]{revtex4}

\usepackage{float}
\usepackage{graphicx}
\usepackage[]{units}
\usepackage{color}

\newcommand{\E}{ \mathbf{E} }
\newcommand{\F}{\mathbf{F}_{\mathrm{ext}}}
\newcommand{\Fe}{\mathbf{F}_{\mathrm{electric}}}
\newcommand{\Fv}{\mathbf{F}_{\mathrm{visc}}}
\newcommand{\Fst}{\mathbf{F}_{\mathrm{stall}}}
\newcommand{\uu}{\mathbf{u}}
\newcommand{\uurel}{\uu_{\rm rel}}
\newcommand{\vv}{\mathbf v (\rr )}

\newcommand{\rr}{\mathbf r}

\newcommand {\be} {\begin{equation}}
\newcommand {\ee} {\end{equation}}
\newcommand {\bea} {\begin{eqnarray}}
\newcommand {\eea} {\end{eqnarray}}

\begin{document}

%%%%%%%%%%%%%%%%%%%%%%%%%%%%%%%%%%%%%%%%%%%%%%%%%%%%%%%%%%%%%%%%%%%%%
\title{Exact expressions for the mobility and  electrophoretic mobility of a weakly charged sphere in a simple electrolyte}
\author{Ludvig Lizana and Alexander Y. Grosberg}
\affiliation{Department of Physics and Center for Soft Matter
Research, New York University, 4 Washington Place, New York, NY
10003, USA}

%%%%%%%%%%%%%%%%%%%%%%%%%%%%%%%%%%%%%%%%%%%%%%%%%%%%%%%%%%%%%%%%%%%%%

% A B S T R A C T

\begin{abstract}
We present (asymptotically) exact expressions for the mobility and electrophoretic mobility of a weakly charged spherical particle in an $1:1$ electrolyte solution.  This is done by analytically solving the  electro and hydrodynamic equations governing the electric potential and fluid flow with respect to an electric field and a
nonelectric force. The resulting formulae are cumbersome, but fully explicit and trivial for computation.  In the case of a very small particle compared to the Debye screening length ($R \ll r_D$) our results reproduce proper limits of the classical Debye and Onsager theories, while in the case of a very large particle ($R \gg r_D$) we recover, both, the non-monotonous charge dependence discovered by Levich (1958) as well as the scaling estimate given by Long, Viovy, and Ajdari (1996), while adding the previously unknown coefficients and corrections. The main applicability condition of our solution is charge smallness in the sense that screening remains linear.
\end{abstract}

\date{\today}

\maketitle

%%%%%%%%%%%%%%%%%%%%%%%%%%%%%%%%%%%%%%%%%%%%%%%%%%%%%%%%%%%%%%%%%%%%%%%%%%%%%%%%%%%%%%%%
%
%                               I N T R O D U C T I O N
%
%%%%%%%%%%%%%%%%%%%%%%%%%%%%%%%%%%%%%%%%%%%%%%%%%%%%%%%%%%%%%%%%%%%%%%%%%%%%%%%%%%%%%%%%
Electrophoresis experiments may be easy to conduct but difficult to
interpret. In a typical setup, an applied field $\E$ moves a charged
particle through a viscous fluid at a steady state velocity $\uu$
such that $\uu=\mu_E \E$, and the particle's electrophoretic
mobility, $\mu_E$, is inferred from measuring $\uu$.
Since the electric field acts not
only on the particle, but also on all other ions in the electrolyte,
which influence the particle's motion, it is generally incorrect to
estimate the electrophoretic mobility as $\mu_E=Q/\gamma$ using Stokes' friction $\gamma =
6\pi \eta R$, where $\eta$ is the fluid viscosity, $R$ is the
particle radius, and $Q$ its charge. The first to develop a comprehensive theory for small ions was Debye and H\"uckel
\cite{debye1923theory}, and subsequently improved by Onsager
\cite{onsager1927theory}, in their famous work on ion conduction.

The extension to the mobility of macroions, such as colloidal
particles or polyelectrolyte molecules  (e.g. DNA), proved to be far
from trivial and has been the subject of numerous works over several
decades \cite{henry1931cataphoresis, booth1950cataphoresis,
levich1962physicochemical, wiersema1966calculation, OBrien_White,
stigter1978electrophoresis, manning1981limiting, teubner1982motion,
cleland1991electrophoretic, russel1992colloidal,
hidalgo1996electrokinetic,long1996simultaneous, overbeek1946smoluchowski,Smoluchowski_electrophoresis, morrison1970electrophoresis} (which by no means is
a complete list).

Traditionally, the electrophoretic mobility is presented using Smoluchowski's formula $\mu_E = \epsilon \zeta/(4\pi \eta)$ \cite{Smoluchowski_electrophoresis}. Here, $\epsilon$ is the dielectric constant of the fluid while $\zeta$ is the electric potential at some ``slip surface'' or ``surface of shear'' pertaining to the macroion. The only case the $\zeta$-potential has an unambiguous meaning, and also where Smoluchowski's relation is exact, is for the weakly charged 
macroion whose surface is so smooth that its local radius of curvature at every point is large compared to the Debye length, $R \gg r_D$ \cite{morrison1970electrophoresis}. In this case $\zeta$ equals the electric potential at the sphere's surface at $R$.
This is, however, not generally  true and the $\zeta$-potential is in fact nothing more than a proxy for electrophoretic mobility. We  demonstrate this explicitly for the charged sphere.

Subsequent works on electrophoresis of colloidal particles can be broadly summarized as forming two groups. One direction was the attempt to solve the cumbersome system of electro-hydrodynamic equations exactly for the spherical particle shape.  Most well known works in this direction are those by Henry \cite{henry1931cataphoresis} and Booth \cite{booth1950cataphoresis} who used perturbative methods.  In the computer era, this line was continued with numerical solutions, most notably by O'Brien and White \cite{OBrien_White}.  The other line of research was presented by Levich \cite{levich1962physicochemical},  Long \textit{et al} \cite{long1996simultaneous}, Morison \cite{morrison1970electrophoresis} and Overbeek \cite{overbeek1946smoluchowski}. These authors did not attempt exact solutions, but provided important physics insights.  We will here use some of these insights to improve and obtain a consistent exact solution for the spherical particle within linear Debye-H\"uckel screening.  At the end, we will return to the comparison of our results to some of the classical ones, especially \cite{henry1931cataphoresis, booth1950cataphoresis}.

Generalizing Smoluchowski theory, Levich \cite{levich1962physicochemical}
applied to a macroion of general shape whose size $R$ is much
larger than the Debye length $r_D$ in the surrounding liquid $R \gg r_D$,
with the idea that in this case every piece of the surface is
approximately planar. Ignoring the fact that electric field is not
necessarily parallel to the surface, Levich predicted $\mu_E$ to be
non-monotonically dependent on charge $Q$, increasing linearly at
small $Q$ but then crossing over to decrease at larger $Q$.
%
% \be \mu_E = \frac{\frac{Q r_D}{4 \pi \eta R^2}}{1 + \frac{3 b
 %r_D^3}{2 R^5 }\frac{Q^2}{k_B T}}
 %\label{eq:Levich_formula_in_our_notation}\ee
%
% with $b$ the microion size in the liquid.
In fact, as Long {\it et al} \cite{long1996simultaneous} explained, the
leading linear in $Q$ behavior in the  $R \gg r_D$ limit can be
established by a simple scaling argument, because in this case the
velocity gradient in the liquid is screened beyond the $r_D$ scale:
balancing the electric force on the macroion $Q\E$ and the drag
force $(\eta R^2/r_D )\uu$, yields $\uu \sim (Qr_D/\eta R^2)\E$ or
$\mu_E \sim Qr_D/\eta R^2$. Based on our exact solution, we will
confirm the correctness of the scaling estimate by Long {\it et al}
\cite{long1996simultaneous}, while correcting the more general
result by Levich \cite{levich1962physicochemical}.

%%%%%%%%%%%%%%%%%%%%%%%%%%%%%%%%%%%%%%%%%%%%%%%%%%%%%%%%%%%%%%%%%%%%%%%%%%%%%%%%%%%%%%%%
%
%                               S Y S T E M   S E T U P
%
%%%%%%%%%%%%%%%%%%%%%%%%%%%%%%%%%%%%%%%%%%%%%%%%%%%%%%%%%%%%%%%%%%%%%%%%%%%%%%%%%%%%%%%%
%%%
% 1. Three inputs
% 2. Force balance = 0
% 3. Accompanying reference frame (stationary motion)
%

In this Letter  we provide (asymptotically) exact expressions for $\mu_E$ and the nonelectric
mobility $\mu_F$ for a weakly charged spherical macroion starting out with the coupled electro
and hydrodynamical equations for the problem. For simplicity we will restrict ourselves to a
two-ion electrolyte (e.g. 1:1 salt). Following the work \cite{long1996simultaneous}, we consider
three independent inputs that control the velocity of the sphere: (i) an electric force $\Fe$
which includes the externally applied field $\E$ and the field due to the surrounding ions, (ii)
a non-electric force $\F$ which could be realized, for instance, by optical tweezers, gravity in particle sedimentation, or by
grafting the colloidal particle to a surface by a polymer, and (iii) a viscous drag force $\Fv$
from the motion of the surrounding liquid.  In steady state, the sphere moves with constant
velocity $\uu$ such that the forces (i)-(iii) are in balance
\begin{equation}\label{eq:forceBal}
\F+\Fe+\Fv = 0  \ .
\end{equation}
Within linear response, the relative velocity of the particle with
respect to the far away wall is
\be \label{eq:mobilities} \uurel = \mu_F \F + \mu_E \E \ ; \ee
(see \cite{todd2011separability} for experimental verification of
this equation).  Therefore, we can find $\mu_F$ and $\mu_E$ by
calculating $\Fe$ and $\Fv$ from the underlying electro and
hydrodynamical equations to linear order in $\uu$ and $\E$. The
problem is known to be technically challenging and we provide, for
the first time, its exact solution. Our derivation relies heavily on
symbolic software (Mathematica) which we believe to be the reason to
why the solution was not obtained a long time ago.

%%%%%%%%%%%%%%%%%%%%%%%%%%%%%%%%%%%%%%%%%%%%%%%%%%%%%%%%%%%%%%%%%%%%%%%%%%%%%%%%%%%%%%%%
%
%                               M A I N  E Q U A T I O N S
%
%%%%%%%%%%%%%%%%%%%%%%%%%%%%%%%%%%%%%%%%%%%%%%%%%%%%%%%%%%%%%%%%%%%%%%%%%%%%%%%%%%%%%%%%

{\it Governing equations.} We formulate the problem in the
accompanying reference frame of the sphere, because there all flows,
currents, and ion distributions are stationary. This means that if
we apply a nonelectric force $\F$ to the particle we must also apply
$-\F$ to the far away walls of the container holding the liquid. In
contrast,  applying an electric field does not require any extra
force applied because the system is overall charge neutral. We
consider therefore the reference frame in which far away walls,
along with the far away liquid, move with velocity $-\uu_{\rm rel}$.
The coupled equations for the hydrodynamics of the fluid, the
electric potential, and ion diffusion are known (e.g.
\cite{OBrien_White}) and summarized below.

%{\it Hydrodynamics.}
The velocity field $\mathbf v$  of the (incompressible) fluid
surrounding the non-slip sphere is described by the low Reynolds
number Navier-Stokes equation
\begin{equation}\label{eq:NS1}
\begin{split} & \eta \Delta \vv - \nabla p(\rr) - \rho(\rr)\nabla
\phi(\rr)=0 \\
& \qquad \mathbf v(R)=0 \ , \quad  \mathbf v (\infty)=-\mathbf
u_{\rm rel} \end{split}
\end{equation}
where $\rho$ is the ion charge density, $\phi$ is the  electric
potential from the surrounding ions and the external electric field
$\E=E\hat z$, $p$ is pressure and $\rr$ designates the vector
distance with respect to the center of the sphere. The  charge
density is related to the ion concentrations $n_\pm$ via  $\rho(\rr)
= e(n_{+}(\rr)-n_{-}(\rr))$; $e$ is elementary charge.

%{\it Electrostatics.}
The electric potential is governed by Poisson equation
\be\label{eq:phi1} \Delta \phi(\rr) = -4\pi \rho(\rr), \ \ \ {\rm
for} \  r>R,
 %
%\Delta \phi(\vec r) &=& 0  , \ {\rm for} \  r<R \nonumber
\ee
where $\Delta \phi_{\rm in}(\rr) = 0  $ inside the sphere; The
dielectric constant is set to unity. We assume that the sphere's
charge is uniformly distributed on it's surface,
%with density $Q/(4\pi R^2)$
leading to standard electrostatic boundary conditions (continuous
tangential component and jump $Q/R^2$ of normal component of
electric field). %The boundary conditions are accordingly
%
%\begin{subequations} \begin{align} \label{eq:phi1BC}
%& \left(\frac{\partial \phi_{\rm in}(\rr)}{\partial r}\right)_{r=R} -
%    \left( \frac{\partial \phi(\rr)}{\partial r}\right)_{r=R}
%     = \frac{Q}{R^2}
%\\ & \left(\frac{\partial \phi_{\rm in}(\rr)}{\partial \theta}\right)_{r=R} -
%      \left(\frac{\partial \phi(\rr)}{\partial \theta}\right)_{r=R} = 0
%\\ & \phi(\rr) \rightarrow -Er\cos\theta \  {\rm as}  \  r \rightarrow
%\infty \ ,
% \ \ \phi_{\rm in}(\rr) < \infty \ \ {\rm as}  \  r \rightarrow 0 \ .
%\end{align}\end{subequations}
%
The ion concentration flux $\mathbf J_\pm(\rr)$ is:
\begin{equation}\label{eq:J1}
\mathbf J_\pm(\rr) = -\mu k_BT \, \nabla  n_\pm(\rr) \mp e\mu n_\pm(\rr)\nabla \phi(\rr) +n_\pm (\rr) \vv
\end{equation}
The first term is the diffusion flux where $\mu$ is ion mobility,
assuming $\mu_{+}=\mu_{-}=\mu$
\footnote{\label{footnote} In principle, $\mu$ is not just the bare mobility
$1/(6\pi\eta b)$; $b$ is ion radius. It must be
corrected due to electric and hydrodynamic forces, the leading
correction being of the order $l_B/r_D $
\cite{LandauLifshitz_Kinetics}. But, keeping the corrected mobility
in Eq. (\ref{eq:J1}) leads to terms $\propto(l_B/r_D)^2$ which is
beyond our linear approximation. We may thus safely put
$\mu=1/(6\pi\eta b)$.},
$k_B T$ is thermal energy. The second term is the drift due to the
electric potential, and the third term is fluid advection. Since the ion flux is stationary and
ions cannot penetrate the sphere's surface,
\begin{equation}\label{eq:J1BC}
\nabla \cdot \mathbf J_\pm(\rr) = 0, \ \ \
\ \
\hat r \cdot  \mathbf J_\pm(\rr)\Big|_{r=R} = 0.
\end{equation}
%

%{\it Calculation of hydrodynamic and electric forces.}
For simplicity we assume that $\E$, $\uu$ and $\F$ are collinear
vectors. The magnitude of $\Fe$ (from Maxwell's stress tensor) and $\Fv$ are thus
\bea\label{eq:dragDef}
&&F_{\rm electric} = \frac Q 2 \int_0^{\pi} \hat r \cdot \left(-\nabla \phi(\rr)\right)_{r=R} \sin \theta d\theta \\
&&F_{\rm visc} = 2\pi R^2\int_0^{\pi}  Ê\left(\sigma_{rr}(R) \cos\theta -
\sigma_{r\theta}(R) \sin \theta \right) \sin\theta d\theta
\nonumber
\eea
%
%where we used that the problem has spherical polar symmetry, and
where $\sigma_{rr}(R)$ and $ \sigma_{r\theta}(R)$ are normal and
tangential elements of the hydrodynamic  stress tensor. The no-slip
boundary condition and the incompressibility of the fluid leads to
$\sigma_{rr}(R) = -p(R)$ and $ \sigma_{r\theta}(R) = \eta
\left(\frac{\partial v_\theta}{\partial r}\right)_{r=R}$.%, where $v_\theta$ is the $\theta$-component of $\mathbf v$.

%%%%%%%%%%%%%%%%%%%%%%%%%%%%%%%%%%%%%%%%%%%%%%%%%%%%%%%%%%%%%%%%%%%%%%%%%%%%%%%%%%%%%%%%
%
%                               M A I N  R E S U L T S
%
%%%%%%%%%%%%%%%%%%%%%%%%%%%%%%%%%%%%%%%%%%%%%%%%%%%%%%%%%%%%%%%%%%%%%%%%%%%%%%%%%%%%%%%%
%
{\it Main results.} We now formulate our results, obtained to  linear order in $\E$ and $\uu$,
postponing to the very end the outline of their derivation.  We find the exact electrophoretic
and nonelectric mobilities to be
\be\label{eq:muE} \mu_E =  \frac{ \frac {Q}{6\pi\eta
R}g\left(\frac{R}{r_D} \right)}{1+ \alpha \left(\frac{Q}{e}
\right)^2
 f\left(\frac{R}{r_D} \right)}, \ \ \mu_F = \frac{\frac
{1}{6\pi\eta R}}{1+ \alpha \left(\frac{Q}{e} \right)^2
f\left(\frac{R}{r_D} \right)} \ee
where $\alpha = b l_B/r_D^2$ with $b$ as ion radius and
$l_B=e^2/(k_BT)$ as Bjerrum length, while $f(R/r_D)$ and $g(R/r_D)$
denote cumbersome but explicit elementary functions (with  ${\rm
Ei}(y)=\int_y^{\infty} (e^{-t}/t) dt$):
\begin{widetext}
\bea \label{eq:fexact} f(y) &=& \frac{e^{2 y} \left[256 \left(2
y^7+6 y^6+6 y^5-45 y^3-45 y^2-45 y+90\right)
   \text{Ei}(2 y)  - 5
\left(y^2-12\right)^2 \left(y^2+3 y+3\right) y^4
   \text{Ei}(y)^2 \right]}{7680 (y+1)^2 \left(y^2+3 y+3\right)}
   \nonumber\\
   &&+\frac{e^y \left(y^9+2
   y^8-22 y^7-51 y^6+12 y^5+150 y^4+1200 y^3+2184 y^2+1152 y-1152\right)
   \text{Ei}(y)}{768 (y+1)^2 \left(y^2+3 y+3\right)}
   \nonumber\\
   &&-\frac{5 y^9+5 y^8+146 y^7+485
   y^6+187 y^5+48 y^4+4692 y^3+10836 y^2-6948 y-3840}{7680 y (y+1)^2 \left(y^2+3
   y+3\right)}
\eea
\end{widetext}
\be\label{eq:Qeff} g(y)=\frac{y^5-y^4-10 y^3+6 y^2 +96 - e^{y}
\left(y^2-12\right) y^4{\rm Ei}(y)}{96 (y+1)}
\ee
The meaning of $g(y)$ is revealed by the fact that effective charge
$Q_{\mathrm{eff}} = Q g(R/r_D)$ is directly measured by looking at
the stall force $\Fst$, which is the amount of $\F$ that must be
applied such that $\uurel=0$: according to equations
(\ref{eq:mobilities}) and (\ref{eq:muE}), $\Fst=-(\mu_E/\mu_F
)\E=-Q_{\rm eff}\E$. The explicit expressions for $f(R/r_D)$
(\ref{eq:fexact}) and $g(R/r_D)$ (\ref{eq:Qeff})
constitute the main achievements of this paper.

\begin{figure}
\includegraphics[width=\columnwidth]{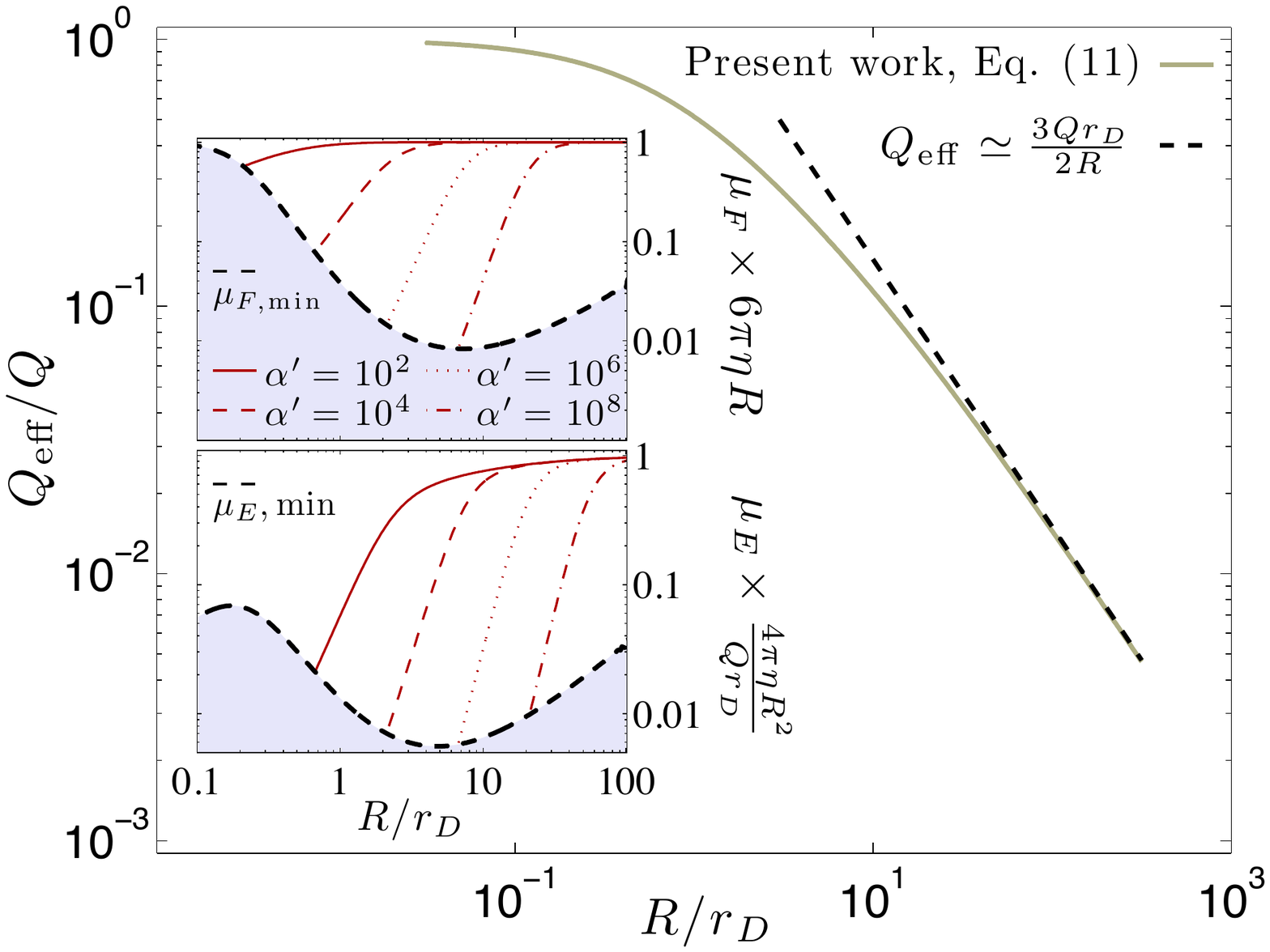}
\caption{(Main panel) Effective charge vs. sphere radius. Solid grey line shows the exact result
(\ref{eq:Qeff}), and the dashed line the leading behavior when $R/r_D\gg 1$. Insets: (upper)
non-electric and (lower) electrophoretic mobility [Eq. (\ref{eq:muE})] for different  $\alpha
(Q/e)^2\equiv \alpha'$. Our theory is applicable in the non-shaded area where the applicability limit lines
$\mu_{F,{\rm min}}$ and $\mu_{E,{\rm min}}$ are Eq. (\ref{eq:muE}) with $Q$ replaced by
$e(R/l_B)^2$.} \label{fig1}
\end{figure}

In Fig. \ref{fig1} (insets) we show $\mu_F$ and $\mu_E$ as functions of particle size for a range
of $\alpha (Q/e)^2$ where we scaled the $y$-axes with the  leading large $R$ behaviors (shown
below). Our model is only valid for a low enough surface charge $Q/R^2\ll e/l_B^2$, and the
shaded areas indicate where this condition is violated. The dividing lines, $\mu_{\rm F,\, min}$
and $\mu_{\rm E,\, min}$, show Eq. (\ref{eq:muE}) where $Q$ was replaced by $e(R/l_B)^2$. The
asymptotic forms read $\mu_{\rm F,\, min}\simeq (6\pi\eta R)^{-1}(1-6\bar \alpha \frac{r_D}{R})$
and $\mu_{\rm E,\, min}\simeq Qr_D(4\pi\eta R^2)^{-1} \times (1-2(2+3\bar \alpha) \frac{r_D}{R})$
for $R/r_D\gg 1$ where $\bar \alpha = r_D^2b/l_B^3$.

The exact expressions for $\mu_E$ and $\mu_F$ are cumbersome but simplify in the limits of large
or small $R/r_D$. When $R/r_D \gg 1$ and $R/r_D \gg \alpha^{1/5} (Q/e)^{2/5}$ we find
\begin{subequations}\begin{align}
& \mu_E \simeq \frac{ Q r_D}{4\pi\eta R^2} \left(1- \frac{4r_D}{R}+\frac{29 r_D^2}{R^2} +...% {\cal O}\left((r_D/R)^3\right)
\right) \label{eq:muE_large_particles_limit}
\\
& \mu_F \simeq \frac{1}{6\pi\eta R}\left(1 - \frac{6\alpha r_D^5}{R^5} \left(\frac Q e\right)^2+ ...%{\cal O}(\alpha (r_D/R)^6)
\right)
\\
& Q_{\rm eff} \simeq
\frac{3Qr_D}{2R}\left(1-\frac{4r_D}{R}+...\right)
\end{align}\end{subequations}
From scaling arguments outlined in the introduction we expect that $\mu_E\sim  Q r_D/(\eta R^2)$
which indeed proves to be the case.
We point out that the same scaling law was obtained earlier (see e.g. \cite{saville1977electrokinetic}) from the argument involving Smoluchowski's formula and the $\zeta$ potential. Imagine that the counter ions are arranged in a spherical shell a distance $r_D \ll R$ away from the sphere, and view this shell as a capacitor.  With capacitance $C\sim R^2/r_D$ and charge $Q$, the potential $\zeta\sim Q /C$ yields the required scaling relation. It is, however, important to realise that the estimate by Long {\it et al} does not involve any assumption of the $\zeta$ potential. Rather it hinges on a balance between electric and hydrodynamic forces which is the very definition of $\mu_E$. Note also that the effective charge decays with increasing bulk
ion concentration $\bar n$ as $Q_{\rm eff} \propto r_D \propto  \bar n^{-1/2}$ (Fig.
\ref{fig1}).

The velocity of the sphere when $\E=0$ is $\uurel =
\mu_F \mathbf F_{\rm ext}$ where, interestingly, $\mu_F$ is {\it not} equal to the Stokes' bare
mobility $1/(6\pi\eta R)$. The reason is that the screening cloud gets deformed as the sphere is
dragged through the electrolyte (or electrolyte flows past the sphere) which gives rise to
relaxation retardation forces. Thus,  $\uurel\sim (6\pi\eta R)^{-1} \F$ is only asymptotically
true (e.g. for a very big sphere, weak surface charge, or a dilute electrolyte $r_D\rightarrow
\infty$). This effect has been addressed elsewhere (e.g. \cite{russel1992colloidal,
barrat1996theory, long1996simultaneous}), in particular in sedimentation literature \cite{booth1954sedimentation, saville1977electrokinetic}. Booth used in \cite{booth1954sedimentation}  the same perturbation scheme as in \cite{booth1950cataphoresis} to calculate the reduction in sedimentation velocity between an uncharged  sphere falling through an electrolyte and a charged sphere screened non-linearly.

Levich \cite{levich1962physicochemical} obtained for large particles ($R\gg r_D$) an expression for
$\mu_E$ (he did not consider $\mu_F$) which has the same type of non-monotonic
$Q$-dependence as our result (\ref{eq:muE}). His result for $g = \frac{3 r_D}{2R}$ is the
correct asymptotics of our result (\ref{eq:Qeff}) at large $R/r_D \gg 1$, but his result for $f =
\frac{3}{2} \left(\frac{r_D}{R} \right)^5$ is different from (\ref{eq:fexact}), thus proving the
necessity to treat properly the curved geometry of the particle surface not parallel to electric
field.

For the opposite limit $\alpha (Q/e)^2 \ll R/r_D\ll 1$ we find
\begin{subequations}\begin{align} \label{eq:muEion} & \mu_E \simeq \frac Q {6\pi \eta R}
\left(1 -  \frac{\alpha r_D}{6 R}\left(\frac Q e\right)^2 - \frac{ R}{r_D} + ...%{\cal O}\left((R/r_D)^2+\alpha\right)
\right)
\\
& \mu_F \simeq \frac{1}{6\pi\eta R} \left( 1-  \frac{\alpha r_D}{6 R}\left(\frac Q e\right)^2 +...%\left(1+{\cal O}(R/r_D)\right)
\right)  \\
&Q_{\rm eff} \simeq Q \left(1-\frac{R}{r_D}+...\right)
\end{align}
\end{subequations}
This result is applicable to a single ion by letting $Q\rightarrow e$ and $R\rightarrow b$
(because $l_B \ll r_D$). The above formula (\ref{eq:muEion}) yields for $\mu_E$ the result
similar but not identical to Onsager's corrected ion mobility for 1:1 salt: $\mu_{E, \rm ion}=
e/(6\pi \eta b)\times[1- l_B/(6(1+1/\sqrt 2 ) r_D) - b/r_D]$ (see e.g.
\cite{LandauLifshitz_Kinetics}, Chp. 10); the only difference is in the numerical coefficient in
the second term. This difference has an interesting physical explanation. In Onsager's
formulation the electrolyte is symmetric in the sense that one type of ion, say  A, is screened
by other ions, B, in the same way as B is screened by A ions. In our case, we consider a solitary
immobile ion which only very weakly contributes to the screening of other ions. Accordingly, if
one takes a more general Onsager model, which involves differences in mobilities of ion species,
then the result reproduces Eq. (\ref{eq:muEion}) in the proper limit exactly, including all
numerical factors (see \cite{LiGro} for details).

In the introduction we pointed out that the $\zeta$-potential, as defined through Smoluchowski's
formula, represents a proxy for the electrophoretic mobility. If we recast our expression for
$\mu_E$ [Eq. (\ref{eq:muE})] onto Smoluchowski's form we identify the $\zeta$-potential for our
problem as $\zeta = (2/3R) Q_{\rm eff}/(1+(Q/e)^2\alpha f)$. Now we may ask, for fixed $Q$ and $R$ what is the ``$\zeta$-surface'', $\rr_\zeta$, on which the potential is equal to $\zeta$:  $\zeta=\phi(\rr_\zeta)$? Is
it close to the sphere's surface? We find that in general it is not, the exception being huge spheres. First, the $\zeta$-surface is not spherically symmetric and 
%Figure \ref{fig:zeta} shows two examples where $\rr_\zeta$  describes a non-spherical
%envelope, its shape depends on $\E$ (and $\uurel$), and
lies somewhere in the bulk outside the sphere. Second, there exists a threshold field $E^*$ (still within applicability of our linear
response theory, $eE^*r_D/k_BT\ll 1$) for which $\rr_\zeta(\theta)$ diverges for some angles $\theta>\theta^*$; in this case the $\zeta$-surface is not even closed at one end. Overall we find that the $\zeta$-surface does not have any simple electric or hydrodynamic interpretation, at least in the case of a spherical and not very big macroion. Additional details is provided in \cite{LiGro}.
%
%\begin{figure}
%\includegraphics[width=\columnwidth]{figure2.pdf}
%\caption{Dashed curves show $\zeta$-potential surfaces, where the electric potential equals the $\zeta$-potential, i.e. $\phi(\rr_\zeta)= \zeta$, in the vicinity of a charged sphere (shaded) for different external electric fields $\bar E = (er_D/k_BT)E$. Note the open envelope for $\bar E=0.1$. Values used: $\alpha (Q/e)^2 = 10^4$, $R/r_D=5$ and $u_{\rm rel}=0$.} \label{fig:zeta}
%\end{figure}
%

%%%%%%%%%%%%%%%%%%%%%%%%%%%%%%%%%%%%%%%%%%%%%%%%%%%%%%%%%%%%%%%%%%%%%%%%%%%%%%%%%%%%%%%%
%
%                                                   D E R I V A T I O N
%
%%%%%%%%%%%%%%%%%%%%%%%%%%%%%%%%%%%%%%%%%%%%%%%%%%%%%%%%%%%%%%%%%%%%%%%%%%%%%%%%%%%%%%%%

{\it Overview of derivation and applicability conditions}. To obtain our results, we  solve the
electro-kinetic equations (\ref{eq:NS1})-(\ref{eq:J1BC}) by linearization. We consider $\E$, $Q$
and $\uu$ as independent control parameters, each sufficiently small so that we may decompose the
electric potential (as well as all other relevant fields such as $\rho(\rr)$, $n_{\pm}(\rr)$, and
$\vv$) as
\be\label{eq:linApprox} \phi(\rr) = \phi_Q(r) + \phi_E(\rr) + \phi_{EQ}(\rr) + \phi_{uQ}(\rr) \ee
The subscripts indicate the expansion terms with respect to the control parameters. For example,
$\phi_Q$ is linear in $Q$ and does not depend on $E$ and $\uu$, whereas $\phi_{EQ}$ is linear in
$E$ and $Q$ and independent of $\uu$, etc.
%
%\footnote{We disregard terms involving $uE$ since the force $QE$ can never be larger than $\eta
%R^2 u$ (how could we otherwise keep the sphere immobile!). At most they can be of the same order
%in which case $uE \sim E^2$.}.
%We restrict ourselves to linear screening and including
%non-linear contributions would add  terms on the order $\phi_{Q^3},\phi_{Q^5}\ldots$ which we
%omit here.
%
The first term in Eq. (\ref{eq:linApprox}) is  screening of the immobile macroion in an
electrolyte. We assume that $\phi_Q$ obeys linear Debye-H\"uckel theory where $e\phi_Q/(k_BT)\ll
1$. For a spherical macroion this means $Q\ll k_BT/e \times R(R+r_D)/r_D$ or, Gouy-Chapman length
$\lambda_G\sim R^2 k_BT/(eQ)$ is much larger than $r_D$. We also assume that the electrolyte
itself obeys linear Debye-H\"uckel theory, that is $l_B/r_D\ll 1$.  Since also $\lambda_G\gg l_B$
the sphere's surface charge must obey  $Q/R^2\ll e/l_B^2$. %This defines our $\E,\uu_{\rm
%rel}\rightarrow 0 $ limit.

The term $\phi_E$ captures polarization effects of the dielectric sphere as well as volume
polarization (induced surface charge due to the deflection of ion currents around the sphere
\cite{dhont2010electric}). The polarization of the ion screening cloud caused by the electric
field is contained in $\phi_{EQ}$. In order to stay within linear response theory
with respect to the electric field we require, in agreement with Onsager
\cite{onsager1996collected}, that $E\ll k_BT/(e r_D)$. That is, the work performed by the
electric field on an elementary charge on the length $r_D$ cannot exceed $k_BT$. The fourth term
includes polarization of the ion cloud due to fluid flow which depends on the velocity of the far
away walls and  $\F$. Following the same logic that set the limit on the electric field leads to
$ F_{\rm ext}\ll k_BT/r_D$. This also holds for the viscous force mediated by the liquid from the
distant walls. As a rough estimate for very low ion concentrations we may put this force to be $
u_{\rm rel} \eta R$ which gives $u_{\rm rel}\ll k_BT/(\eta Rr_D)$.  Similar interpretation exists
also for expansion terms of other relevant fields.

Thus, our calculation is exact in first perturbation order with respect to electric field $\E$
and velocity $\uu$.  Nevertheless, it correctly captures the very non-linear $Q$-dependence in
both $\mu_E$ and $\mu_F$ (\ref{eq:muE}).  This arises from the electric force
(\ref{eq:dragDef}) term proportional to $Q \phi_{uQ}$.

{\it Concluding remarks.} One of the most surprising aspects of our results is that we consider linear Debye-H\"uckel screening and obtain the mobility Eq. (\ref{eq:muE}) which is decidedly non-linear in particle charge $Q$.  How is it possible?  As an example consider the case when Bjerrum length $l_B$ is much smaller than relevant ion size $b$ (related to the mobility $(6\pi \eta b)^{-1}$, i.e., including the solvation shell), $l_B \ll b$, then there is parametric range of particle sizes $R$ such that $1 \gg R/r_D \gg l_B/b$, and then there is a range of particle charge $Q$ such that $(R r_D/b l_B)^{1/2} \ll Q/e \ll R/l_B$.  The right inequality assures that screening remains linear, while the left inequality guarantees domination of the second term in the denominator of our equation (\ref{eq:muE}), leading to mobility dropping of as $1/Q$ -- very non-linear in $Q$. This example is not the only one, and possibly not even the most interesting one. But the point is, linearity of screening and linear dependence of mobility on charge are controlled independently from one another. From the more qualitative physics view point, what we (as well as our predecessors \cite{Smoluchowski_electrophoresis, henry1931cataphoresis, overbeek1946smoluchowski, booth1950cataphoresis, booth1954sedimentation, levich1962physicochemical, morrison1970electrophoresis, OBrien_White, long1996simultaneous}) consider is a linear response theory.  As such, it treats particle velocity as linear in both applied electric field and applied non-electric force (as was powerfully emphasized by Long {\it et al.} \cite{long1996simultaneous}).  There is no general physics principle saying that linear response should be linear in particle charge. The inspection of our solution indicates that non-linear dependence of mobility on charge descends from convective distortion of the ion cloud and, therefore, can exist when screening is linear.

At the end, we are now in a position to compare our results to the classical ones by Henry \cite{henry1931cataphoresis} and Booth \cite{booth1950cataphoresis}.  Henry \cite{henry1931cataphoresis} considered ``cataphoresis'' (which is presumably how electrophoresis was called at the time) of a charged sphere as early as 1931.  The major approximation employed in his approach was the assumption that the ion cloud maintains its spherical symmetry. This means, in our notation,  that the electric force on the particle is $\E Q$, the convective force from the liquid is Stokes' drag $-6\pi\eta R \uurel$, and the electro-osmotic drag force comes from the body force $-\rho_Q\E$ in Navier-Stokes equation, $\rho_Q$ being the spherically symmetric equilibrium charge density. Balancing these forces leads to Henry's electrophoretic mobility.
From Onsager's work on small ions \cite{onsager1927theory} we know that because of important retardation forces (also known as the relaxation effect) the validity of this assumption is limited to the case of thin Debye layers ($R/r_D \gg 1)$.  Indeed, in this limit our result (\ref{eq:muE_large_particles_limit}) coincides with that of Henry.  Of course, this is not so surprising in the light of scaling argument due to Long \textit{et al} \cite{long1996simultaneous}.  More generally, the contribution from ion cloud relaxation in our treatment is the $\sim Q^2f(y)$ term in the denominator of Eq. (\ref{eq:muE}); in other words, relaxation effect is exactly the reason why the denominator in Eq (\ref{eq:muE}) is \textit{not} unity.  If we formally replace this denominator with unity, we obtain exactly the result by Henry: our $g(y)$ [Eq. (\ref{eq:Qeff})] except for notations, is exactly the expression one obtains by plugging in the Debye-H\"uckel potential ($\phi_Q$) into Henry's Eq. (18).

In a two decades later study \cite{booth1950cataphoresis} Booth did include the relaxation effect as well as deformation of the ion cloud due to Stokes drag; The convective flow acts not only on the sphere when it moves through the liquid but also on the surrounding counter- and co-ions.  He also included non-linear screening on the level of mean-field Poisson-Boltzmann equation. Booth's calculation is based on a series expansion  in $Q$ where he expresses the  steady-state electrophoretic velocity as the polynomial $U = \sum_{\nu=1}^\infty c_\nu Q^\nu$  and seeks $c_\nu$. He recovered Henry's mobility in $c_1$ and found, as we do, that $c_2=0$ (which is in fact true for all even indices $c_2=c_4=...=0$ for obvious symmetry reasons). However, due to the mathematical complexity of the problem Booth was only able to get numerical values for  $c_{\nu}$ for $ \nu \geq 3$ from  complicated integral expressions.  In our work we took a different path: we adopted the clever observation by Long {\it et al} \cite{long1996simultaneous} that the sphere's velocity is a linear combination of $\F$ and $\E$, and in this way derive an \textit{exact} expression for $\mu_E$ (and $\mu_F$), not in terms of a formal power series. Unlike Booth we stayed within Debye-H\"uckel linear screening and found, in agreement  with Levich \cite{levich1962physicochemical}, that $\mu_E$ depends non-monotonically on $Q$. This means that the non-linearity in $\mu_E$ is \textit{not} the consequence of non-linear screening but exists also in the linear screening regime and stems from convective distortion of the ion cloud. Of course, including the non-linear screening effects as well as beyond mean field correlation effects will correct pre factors to $Q^3$, $Q^5$ etc., in our analysis but our work put the electrophoresis of a Debye-H\"uckel sphere on firm ground, and Eq. (\ref{eq:muE}) is thus new. This constitutes the novelty of our work.
%
%This is an important ingredient in our analysis and comes out ``automatically''  since we adopted the clever observation by Long et al \cite{long1996simultaneous} that the sphere's velocity is a linear combination of $\F$ and $\E$ terms.  Therefore, Booth's result for $\mu_E$ should agree with our's only to linear order in $Q$ -- which it does.   The non-linear dependence on $Q$, which leads to non-monotonic dependence on $Q$, arises from relaxation effects (considered by Booth) and convective flow (neglected by Booth), just like in Levich's work \cite{levich1962physicochemical}.   The conclusion is that our formula (\ref{eq:muE}) is a novel result which is exact to within Debye-H\"uckel screening approximation.
%
In the future, it might be desirable to consider a more complete theory which must go not only beyond linear Debye-H\"uckel screening, but also beyond the non-linear mean field Poisson-Boltzmann equation. In this setting completely new phenomena arise such as charge inversion, like charge attraction etc. \cite{grosberg2002colloquium}.  To include these type of phenomena in the theory of electrophoresis is a real challenge.

{\it Summary and outlook}.  In this Letter, we presented an asymptotically exact solution for the
electrophoretic mobility for a weakly charged colloidal sphere. We established that the heavily
used   $\mu_E \sim Qr_D/\eta R^2$ is asymptotically correct in the linear approximation,
and that the non-monotonous dependence on $Q$ of $\mu_E$ and $\mu_F$ is a consequence of the system's dynamics rather than
non-linear screening.  We assumed linear
screening, no slip hydrodynamics, laminar fluid flow, equal dielectric constants of the particle and
solvent, and ideal spherical shape which we consider as acceptable restrictions to produce a
solution which is exact.
Our expression for $\mu_E$ (and $\mu_F$) is analogous to the
appealingly simple Stokes' formula $1/(6 \pi \eta R)$. Stoke's
formula has applicability limitations (e.g. spherical shape and no
slip) but is nevertheless enormously useful. Our formulas are more
complicated and significantly more restricted since the problem is
much more complex. But we hope that, being exact, they will find their
usefulness in  a range of applications.

We thank Andrew Hollingsworth for pointing out some useful references and Paul Chaikin for
encouragement. We gratefully acknowledge the hospitality of KITP, Santa Barbara, where part of
this work was performed. This research was supported in part by the National Science Foundation
under grant no. NSF PHY11-25915.  LL acknowledges financial support from the Knut and Alice
Wallenberg foundation. Finally we wish to thank the anonymous referee for great help provided by his/hers critical remarks.

%%%%%%%%%%%%%%%%%%%%%%%%%%%%%%%%%%%%%%%%%%%%%%%%%%%%%%%%%%%%%%%%%%%%%%%%%%%%%%%%%%%%%%
% B I B L I O G R A P H Y

\end{document}